\documentstyle[preprint,aps]{revtex}
\begin{document}
\draft

\title{Using binary stars to bound the mass of the graviton}

\author{Shane L.\ Larson\cite{Lar} and William A.\ Hiscock\cite{His}}
\address{Department of Physics, Montana State University, Bozeman,\\
Montana 59717}

\date{December 22, 1999}

\preprint{MSUPHY99.08}

\maketitle

\begin{abstract}
Interacting white dwarf binary star systems, including helium
cataclysmic variable (HeCV) systems, are expected to be strong sources
of gravitational radiation, and should be detectable by proposed
space-based laser interferometer gravitational wave observatories such
as LISA.  Several HeCV star systems are presently known and can be
studied optically, which will allow electromagnetic and gravitational
wave observations to be correlated.  Comparisons of the phases of a
gravitational wave signal and the orbital light curve from an
interacting binary white dwarf star system can be used to bound the
mass of the graviton.  Observations of typical HeCV systems by LISA
could potentially yield an upper bound on the inverse mass of the
graviton as strong as $h/m_{g} = \lambda_{g} > 1 \times 10^{15}$ km
($m_{g} < 1 \times 10^{-24}$ eV), more than two orders of magnitude
better than present solar system derived bounds.
\end{abstract}


\section{INTRODUCTION}\label{sec:intro}

The advent of large scale laser interferometer gravitational wave
detectors promises to lead to the direct detection of gravitational
radiation, which will create a new observational field of science:
gravitational wave astronomy. The introduction of gravitational waves
into the retinue of astronomical observations will yield
important new information about the dynamics of astrophysical
systems, and will also provide an excellent opportunity to conduct
new tests of gravity.

Proposed space-based laser interferometers, such as the LISA (Laser 
Interferometer Space Antenna) \cite{LISA} and OMEGA (Orbiting Medium 
Explorer for Gravitational Astrophysics) \cite{OMEGA} observatories, 
will be particularly well poised to begin astrophysical studies since 
there are known sources of gravitational radiation which will be 
easily visible to these instruments, namely interacting binary white 
dwarf (IBWD) star systems.

The IBWD sources are particularly appealing targets because they are 
good candidates for simultaneous optical and gravitational wave 
observations.  Many of the IBWDs which are known to exist are being 
studied and monitored by the Center for Backyard Astrophysics 
(CBA)\footnote{The CBA is a network of amateur astronomers equipped 
with CCD photometry equipment who monitor variable stars.  The network 
is managed by professional astronomers at Columbia University.}, and 
are expected to be strong sources of monochromatic gravitational waves 
which should be easily visible to an instrument such as LISA with only 
a few minutes of signal integration.  Simultaneous optical and 
gravitational wave observations will be useful in refining the current 
physical models used to describe these systems, and for testing 
relativistic theories of gravity in the radiative regime by comparing 
the propagation speeds of electromagnetic and gravitational wave 
signals.

This paper examines how the comparison of the phase of the orbitally 
modulated electromagnetic signal (the light curve) and a gravitational 
wave signal from an IBWD star system can be used to bound the mass of 
the graviton.  If the mass of the graviton is assumed to be known by 
other measurements, then the observations may be used to determine the 
properties of the binary star system being monitored.

Current conservative bounds on the graviton mass come from looking
for violations of Newtonian gravity in surveys of planetary motions
in the solar system.  If gravity were described by a massive field,
the Newtonian potential would have Yukawa modifications of the form
\begin{equation}
    V(r) = -{M \over r}\exp(-r/\lambda_{g})
    \label{YukawaPotential}
\end{equation}
where $M$ is the mass of the source of the potential, and
$\lambda_{g} = h/m_{g}$ is the Compton wavelength of the graviton,
where $m_{g}$ is the graviton mass. The current best bound on the graviton
mass from planetary motion surveys is obtained by using Kepler's
third law to compare the orbits of Earth and Mars, yielding
$\lambda_{g} > 2.8 \times 10^{12}$ km ($m_{g} < 4.4 \times 10^{-22}$
eV) \cite{Talmadge}.

Another bound on the graviton mass can be established by considering
the motions of galaxies in bound clusters \cite{Goldhaber74},
yielding $\lambda_{g} > 6 \times 10^{19}$ km ($m_{g} < 2 \times
10^{-29}$ eV). This bound, while stronger than solar system
estimates, is considerably less robust, due to uncertainty about the
matter content of the Universe on large scales ({\it e.g.}, the
amount and nature of dark matter is widely debated, and uncertain at
best).

Recent work by Will \cite{WillGraviton} has suggested that the
mass of the graviton could be bounded using gravitational wave
observations.  If the graviton is a massive particle, then the
speed of propagation of a gravitational wave will depend on its
frequency.  As binary systems evolve, they will slowly spiral
together due to the emission of gravitational radiation.  Over
the course of time, the frequency of the binary orbit rises,
ramping up rapidly in the late stages of the evolution, just
prior to coalescence.  Laser interferometer gravitational wave
detectors should be able to track the binary system's evolution,
obtaining the detailed time-dependent waveform using the matched
filtering techniques required for data analysis in these
detectors.

Space-based detectors such as LISA will be able to observe the
coalescence of massive ($\sim 10^{5}$ to $\sim 10^{7}
\mbox{M}_{\odot}$) binary black holes, as well as the gravitational
wave emission from compact binary star systems which are far from
coalescence ({\it e.g.}, interacting binary white dwarfs).  Ground-based
detectors such as LIGO will be able to detect the merger of smaller
black hole binaries ($\sim 10 \mbox{M}_{\odot}$), as well as the
coalescence of compact binary stars ({\it e.g.}, neutron star/neutron
star binaries).  If the graviton is a massive particle, then the
observed signal will not perfectly match theoretical templates
computed using general relativity theory, in which the graviton is
massless; a massive graviton would cause dispersion in the
gravitational waves. By using matched filtering of inspiral
waveforms, this dispersion could be bounded, thereby bounding the
mass of the graviton. Will \cite{WillGraviton} finds that LIGO could
bound the graviton mass at $\lambda_{g} > 6.0 \times 10^{12}$ km
($m_{g} < 2.1 \times 10^{-22}$ eV) by observation of the inspiral of
two $10\ \mbox{M}_{\odot}$ black holes.  A space-based interferometer
such as LISA, observing the inspiral of two $10^{7}\
\mbox{M}_{\odot}$ black holes could bound the graviton mass at
$\lambda_{g} > 6.9 \times 10^{16}$ km ($m_{g} < 1.8 \times 10^{-26}$
eV). If the graviton is massive, then these numbers represent the
minimum masses detectable by such observations.

The analysis in this paper shows that LISA observations of known
IBWD sources could yield a bound as strong as $\lambda_{g} > 5
\times 10^{15}$ km ($m_{g} < 2 \times 10^{-25}$ eV),
considerably stronger than present solar system based bounds. The
IBWD bound also has the advantage of not depending on the
complicated details of black hole coalescences.

Section \ref{sec:IBWD} reviews what is known about the interacting 
binary white dwarfs, in particular the helium cataclysmic variable 
(HeCV) systems and their archetype, the binary AM CVn (AM {\it Canum 
Venaticorum}).  Section \ref{sec:GMEM} reviews the basic notions associated 
with (possibly) massive photons and gravitons.  Sections 
\ref{sec:GravitonMass}$-$\ref{sec:PhaseDelay} propose a new experiment 
to measure the graviton mass using IBWD observations, and an 
expression for the mass is derived.  In Section \ref{sec:AMCVn}, the 
sensitivity predicted for LISA is used to estimate how precise a bound 
could be placed on the graviton mass from HeCV observations.  Section 
\ref{sec:Ending} summarizes the results, and also suggests how 
correlation of phase measurements might be used to measure other 
astrophysical parameters in the binary system, such as the accretion 
disk radius.

Throughout this paper, geometric units with $G = c = 1$ are employed, 
unless otherwise noted.

\section{Interacting Binary White Dwarf Systems}\label{sec:IBWD}

Estimates suggest the Galaxy is populated by $\sim 10^{7}$ close 
binary star systems \cite{HilsWebbink}.  The sheer numbers of these 
systems is likely to have profound consequences for space-based 
gravitational wave observatories.  The combined gravitational waves 
from these binaries will produce a stochastic background which rises 
well above the low frequency detection sensitivity of LISA 
\cite{HilsBender} .  Particularly strong ({\it e.g.}, nearby) binary 
systems will rise above this background and be observable by a 
spaceborne observatory.  One class of such sources are the helium 
cataclysmic variable (HeCV) stars.  The properties of the six nearest 
known HeCVs are shown in Table \ref{BinaryTable} 
\cite{HellingsPrivate}.  The predicted stochastic gravitational wave 
background due to short period binary stars, as calculated by Hils and 
Bender \cite{HilsBender} and the predicted signals of the six nearest 
HeCVs are plotted in Figure \ref{WDCurve}, along with the predicted 
sensitivity curve of LISA. If present models of the spatial density of 
close binaries in the Galaxy are correct, roughly 5000 of these 
sources should be individually detectable by a space-based laser 
interferometer such as LISA\cite{LISA}.

Currently the best models for HeCVs describe a star system where the
secondary star (the lower mass companion in the binary system, usually
a degenerate helium dwarf star) has expanded to fill its Roche lobe,
and the primary star (the larger mass, compact white dwarf) lies at
the core of an accretion disk.  Matter overflows from the secondary
Roche lobe and streams onto the accretion disk, creating a hot spot
which emits a strong electromagnetic signal.

There are several mechanisms whereby the light curve of an HeCV could 
be modulated as seen from the perspective of observers on Earth.  The 
simplest model is for systems whose orbital plane is close to the line 
of sight to the Earth, so that the stars periodically eclipse, 
partially or wholly.  The eclipse phase could dim the stellar 
components, part of the primary emission regions of the accretion 
disk, or the Roche lobe.  Another possible mechanism for variation in 
the light curve is associated with the emission from the secondary 
star, which has expanded to fill its Roche lobe.  The star will appear 
brighter when the largest surface area is presented to the observer 
along the projected line of sight.  This will occur twice in each 
orbit, when the line of sight is perpendicular to the line of centers.  
Signatures in the light curve due to projected area effects are called 
``ellipsoidal variations,'' and are most easily observed in infrared 
wavelengths \cite{WadeWard}.  Yet another possible mechanism is 
currently favored to explain the source of variation in the light 
curve of the archetype system for HeCVs, AM CVn \cite{AMCVnPeriod}.  In this 
model the hot spot on the accretion disk radiates approximately 
radially outward from the disk.  As the binary orbits, this hot spot 
alternately turns towards and away from distant observers, leading to 
a modulation of the light curve (a so-called ``flashlight'' 
mechanism).

A detailed theoretical model of AM CVn has been constructed,
describing a variety of signals which are present in the
photometric data \cite{Faulkner1972,Patterson1992,Patterson1993}.
This model suggests that AM CVn is a member of a class of
variable stars that have periodic features in the light curve
known as ``superhumps'' \cite{Vogt1982,Whitehurst88}.  The model
explains the  superhump feature as being caused by the existence
of an eccentric precessing accretion disk, with a precession
period which is slightly longer than the orbital period of the
binary. Knowing the superhump period, $P_{sh}$, and the
precessional period of the accretion disk (apsidal advance),
$P_{aa}$, the model predicts the {\it orbital} period will be
given by
\begin{equation}
    P_{orb}^{-1} =  P_{sh}^{-1} +  P_{aa}^{-1} \ .
    \label{PeriodTheory}
\end{equation}
Photometry of AM CVn shows the existence of a superhump signature
at $P_{sh} = 1051.2$ s, and the period of the accretion disk
precession at $P_{aa} = 13.38$ hr.  Using Eq.\
(\ref{PeriodTheory}) this model predicts a binary orbital period
of $P_{orb} = 1028.77 \pm 0.18$ s for AM CVn
\cite{Patterson1993}.  Photometric observations by the CBA have
recently confirmed an orbital period of $P_{orb} = 1028.7325 \pm
0.0004$ s \cite{AMCVnPeriod}.

\section{ Massive Gravitons and Photons}\label{sec:GMEM}

To constrain the mass of the graviton by comparing the propagation 
speed of gravitational and electromagnetic waves, one must consider 
how the speed of gravitational waves (and electromagnetic waves) is 
related to the mass of the graviton (and the possible mass of the 
photon).

The current bound on the photon mass is $m_{\gamma} < 2 \times
10^{-16}$ eV \cite{PDG}, which is much larger than the current
bounds on the graviton mass ({\it cf.}, the solar system bound
on the graviton mass is $m_{g} < 4.4 \times 10^{-22}$ eV).  Is it
then justifiable to treat the photon as a massless particle,
while at the same time treating the graviton as a massive
particle?

The resolution to this question can be understood by examining
the partition of energy between the rest mass and kinetic energy
of a particle being received from a distant binary.  From the
relativistic energy, $E^{2} = p^{2} + m^{2}$, one may write the
velocity of any particle as
\begin{equation}
   v^{2} = 1 - {m^{2} \over E^{2}}\ .
   \label{RelVelocity}
\end{equation}
If $m \ll E$, then Eq.\ (\ref{RelVelocity}) implies
\begin{equation}
   \varepsilon = 1 - v \simeq {1 \over 2}{m^{2} \over E^{2}} \ ,
   \label{VelocityDiff}
\end{equation}
where $\varepsilon$ parameterizes the difference between the
velocity of the particle and $c$.

For optical photons ($\lambda \sim 500$ nm) received from a
binary star system, the characteristic energy is $E_{\gamma}
\simeq 2.5$ eV. For this energy, $\varepsilon_{\gamma} \leq 3
\times 10^{-33}$. Similar considerations may be applied to the
gravitons received from the same binary systems.  In this case,
the frequency of the gravitational waves is $f \sim 10^{-3}$ Hz,
giving a characteristic energy for a single graviton of $E_{g}
\sim 4 \times 10^{-18}$ eV. Using the solar system bound on the
graviton mass, $m_{g} < 4.4 \times 10^{-22}$ eV, yields
$\varepsilon_{g} \leq 1 \times 10^{-8}$.

For the current bounds on the photon and graviton masses,
$\varepsilon_{\gamma} \ll \varepsilon_{g}$.  If the bound on the
mass of the graviton is not drastically improved ({\it e.g.},
decreasing the bound on $m_{g}$ by $12$ orders of magnitude, such
that $\varepsilon_{g} \sim \varepsilon_{\gamma}$), then the
effect of a non-zero mass will be much more significant for
gravitons than photons in our analysis. This justifies the
treatment of the photons as massless particles and the gravitons
as massive particles in this paper.

Since the gravitational waves emitted by an HeCV binary star
system are essentially monochromatic, they will have a single
particular velocity,
\begin{equation}
     v_{g} \simeq 1 - {1 \over 2} \left( {m _{g} \over E} \right)^{2} \ .
     \label{vGraviton}
\end{equation}
Writing the energy in Eq.\ (\ref{vGraviton}) as $E = \hbar
\omega$, and identifying the Compton wavelength of the graviton
as $\lambda_{g} = h/m_{g}$ this becomes
\begin{equation}
    v_{g} \simeq 1 - {1 \over 2} \left( { {2 \pi} \over {\lambda_{g}
    \omega}} \right)^{2} \ .
    \label{vGraviton2}
\end{equation}

\section{Correlation of Electromagnetic and Gravitational Observations
to Measure the Graviton Mass}\label{sec:GravitonMass}

As noted in Section \ref{sec:IBWD}, it is expected that the
Galaxy harbors a large population of interacting binary white dwarf
stars. The light curves for several of these systems are
already known, obtained from ground-based optical photometry.
Because these systems are expected to be observable in the
gravitational wave spectrum (in addition to the optical
spectrum), they present an excellent opportunity to directly
compare the propagation speed of electromagnetic and
gravitational waves.

Consider the schematic diagram shown in Figure \ref{BinaryIdea}.  The 
phase fronts of the light curve modulation are represented in the top 
half of the diagram.  The binary star system will also emit 
gravitational radiation which could be monitored by Earth-bound 
observers as well.  The phase fronts of the gravitational wave signal 
are represented in the lower half of the diagram\footnote{Assuming a 
circular orbit, the frequency of the gravitational radiation is 
actually twice the orbital frequency of the binary.  For clarity, only 
half the gravitational phase fronts have been drawn in the diagram.}.

Suppose the two signals are emitted in phase at the source as
shown. If the graviton is a massive particle, then the
gravitational waves propagate at a speed $v_{g} < 1$, and the
gravitational phase fronts will {\it lag} behind the light curve
phase fronts when the signals arrive at Earth, as shown. By
measuring the lag between the phase fronts, the mass of the
graviton can be measured or bounded.

To determine the lag between the two signals, the phase of each signal 
must be measured.  Consider a binary with orbital frequency 
$\omega_{o}$ at a distance $D$ from Earth.  Assuming the photon to be 
massless ($v = c = 1$), the observed phase of the light curve will be
\begin{equation}
    \phi_{em} = \omega_{o}D + A \ ,
    \label{EMPhase}
\end{equation}
where the term $A$ represents a variety of effects (discussed in
Section \ref{sec:PhaseDelay}) which could create phase delays
between the electromagnetic and gravitational signals that are
being monitored. In contrast, the gravitational wave signal,
traveling at $v_{g} < 1$ will arrive at Earth with a phase of
\begin{equation}
    \phi_{gw} = 2\omega_{o}{D \over v_{g}} \ .
    \label{GWPhase}
\end{equation}

The phase lag, $\Delta$, between the light curve and
gravitational wave signals is constructed from these two phases:
\begin{eqnarray}
    \Delta & = & {1 \over 2}\phi_{gw} - \phi_{em} \nonumber \\
    & = & \omega_{o}D \left( {1 \over v_{g}} - 1 + \alpha \right) \ ,
    \label{PhaseLag}
\end{eqnarray}
where $\alpha = A/(\omega_{o}D)$.  The factor of $1/2$ insures that
the phase subtraction is done between two signals with the {\it same}
frequency.  It is convenient to define the fractional change in the
phase as
\begin{equation}
    \epsilon = {\Delta \over {\omega_{o}D}} = \left( {1 \over v_{g}} -
    1 + \alpha \right) \ .
    \label{FractionalLag}
\end{equation}

Taking the definition of $v_{g}$ from Eq.\ (\ref{vGraviton2}) and
substituting into Eq.\ (\ref{FractionalLag}), the Compton
wavelength of the graviton as a function of the fractional phase
lag is found to be
\begin{equation}
    \lambda_{g} = { \pi \over \omega_{o}}\sqrt{{1 \over 2}
    \left ( 1 + {1 \over {\epsilon - \alpha}}\right )} \ .
    \label{ComptonWavelength}
\end{equation}

An obvious question to ask about this analysis is how to determine
whether or not the phase difference between the two signals is greater
than a single cycle, and hence undetectably large ({\it e.g.}, $\Delta
= 2 n \pi + \xi$, where $\xi$ is a small quantity).  One can eliminate
this concern because strong bounds on the graviton mass already exist.
The largest observable phase shift which is consistent with current
bounds can be computed by simply evaluating Eq.\
(\ref{ComptonWavelength}) with $\lambda_{g}$ equal to the bound of
interest.  For example, the bound on the graviton mass given by solar
system constraints, applied to the AM CVn system, yields a maximum
fractional phase change of $\epsilon = 1.5 \times 10^{-9}$.  Since AM
CVn lies at a distance of $D = 101$ pc, this value of $\epsilon$
indicates a maximum phase difference of $\Delta = 9.6 \times 10^{-2}$
for that system (we have let $\alpha \rightarrow 0$ here for
convenience; if the measured phase difference were larger than this
value, it would indicate that $\alpha \neq 0$).

\section{Phase Delays}\label{sec:PhaseDelay}

In order to evaluate Eq.\ (\ref{ComptonWavelength}), one must not only
measure the phase lag between the two signals, but an estimate must be
made for the value of $\alpha$.

The parameter $\alpha$ can be written as the sum of two primary
sources of delay between the gravitational and electromagnetic signal
phases:
\begin{equation}
     \alpha = \alpha_{\star} + \alpha_{\rm path} \ ,
     \label{AlphaDefine}
\end{equation}
where $\alpha_{\rm path}$ is a phase lag associated with the wave's 
propagation from the binary to the observer at the Earth and 
$\alpha_{\star}$ is a phase lag which depends on the specific 
astrophysical nature of the binary star system.

In principle, $\alpha_{\rm path}$ will be nonzero because the line of 
sight to the binary is an imperfect vacuum, with non-unit index of 
refraction.  The variations in index of refraction over the path will 
cause a lag in the electromagnetic signal.  The dominant source of 
this lag will be caused by propagation of the signal through the 
Earth's atmosphere.

A simple estimate of the value of $\alpha_{\rm path}$ can be made
by computing the electromagnetic phase delay due to propagation
through a modeled exponential atmosphere, with a density profile
\begin{equation}
    \rho(r) = \rho_{o}\exp(-r/h_{s}) \ ,
    \label{DensityProfile}
\end{equation}
where $h_{s}$ is the scale height of the atmosphere.  If the index of
refraction, $n$, is assumed to vary linearly with the density, then
\begin{equation}
    n(r) = 1 + \eta \exp(-r/h_{s}) \ ,
    \label{IndexRefraction}
\end{equation}
where $\eta = (n_{atm} - n_{vac}) = (n_{atm} - 1)$ is the
difference in index of refraction between the atmosphere and
vacuum.  The index of refraction is related to the signal's propagation
speed $v$ by
\begin{equation}
    v = {dr \over dt} = {1 \over n(r)} \ .
    \label{PropagationSpeed}
\end{equation}
Eq.\ (\ref{PropagationSpeed}) can be integrated using Eq.\
(\ref{IndexRefraction}) to obtain
\begin{equation}
    \int_{0}^{t_{transit}} dt = \int_{0}^{r_{o}} dr \left[ 1 + \eta
    \exp(-r/h_{s}) \right] \ ,
    \label{indexIntegral}
\end{equation}
where $t_{transit}$ is the time it takes a photon to transit the
atmosphere, and $r_{o}$ is the height at which the effects of the
atmosphere become negligible.  Completing the integration yields
\begin{equation}
    t_{transit} = r_{o} + h_{s} \eta(1 - e^{-r_{o}/h_{s}}) \ .
    \label{TimeTransit}
\end{equation}
In order to compute a phase delay, one is interested in the time by
which the photons are {\it delayed} by the atmosphere, which is
\begin{eqnarray}
    t_{delay} & = & t_{transit} - t_{vacuum} \nonumber \\
    & = & t_{transit} - r_{o} \ ,
    \label{TimeDelay}
\end{eqnarray}
where $t_{vacuum}$ is the time it would take a photon to travel the
same distance ({\it i.e.}, the atmospheric depth) in vacuum.

The phase delay introduced by this effect is obtained by multiplying
the delay time by the frequency of the signal being observed, 
\begin{equation}
    \phi_{delay} = \omega_{o} t_{delay} \ .
    \label{PhaseDelay}
\end{equation}
The parameter $\alpha_{path}$ is obtained from the phase delay by
dividing by the reference phase, $\omega_{o} D$ (as in Eq.\
(\ref{PhaseLag})), giving
\begin{equation}
    \alpha_{path} = {\phi_{delay} \over {\omega_{o} D}} = {t_{delay}
    \over D} \ .
    \label{AlphaATM}
\end{equation}

It is now possible to numerically estimate the delay introduced by the
atmosphere.  Assuming the Earth's atmosphere to be in hydrostatic
equilibrium gives a scale height $h_{s} = 8500$ m.  A typical value
for the difference between the atmosphere's index of refraction and
unity, evaluated at sea level, is $2.8 \times 10^{-4}$.  Taking the
atmospheric depth to be $r_{o} = 10 h_{s}$, Eq.\ (\ref{TimeDelay})
gives $t_{delay} = 8.0 \times 10^{-9}$ s.  For the prototypical source
AM CVn, at a distance $D = 101$ pc, Eq.\ (\ref{AlphaATM}) yields
$\alpha_{\rm path} = 7.7 \times 10^{-19}$.  This is much smaller than
any possible phase measurement, either electromagnetic or
gravitational, and so we will henceforth ignore this source of
uncertainty.

The parameter $\alpha_{\star}$ is a measure of the initial
phase difference at the source between the gravitational wave and
electromagnetic signals. It indicates the relative phase
difference between the peaks in the light curve of the binary,
and the peaks in the quadrupole gravitational radiation pattern.
Determining the value of $\alpha_{\star}$ requires knowledge of
the position of the stars in the binary system when the
electromagnetic signal peaks.  The quadrupole gravitational
radiation pattern will peak along the line of masses in the
binary system, and also $180^{\circ}$ away from the line of masses
(since the frequency of the gravitational radiation is
$\omega = 2\omega_{o}$).  If the primary variation in the
binary light curve is associated with the transit of the line of
masses across the observer's line of sight ({\it e.g.}, the
system is an eclipsing binary), then it is straightforward to
assign $\alpha_{\star} = 0$, indicating no initial phase delay between
the gravitational wave signal and the binary's light curve. 

For more complicated systems such as AM CVn, the light curve variation
reflects the orbital motion of the hot spot on the edge of the
accretion disk where the matter stream from the secondary Roche lobe
overflow strikes the disk.  Studies of Roche lobe overflow
\cite{Flannery75} show that the transferred material remains in a
coherent matter stream which spirals in towards the primary star.  In
this sort of system, the location of the hot spot will determine the
value of $\alpha_{\star}$, which will describe the amount by which the
hot spot leads the line of masses, as shown in Figure \ref{RocheLobe}.

Estimates of the size of the primary accretion disk in HeCV type
systems suggest that disk radii will be around $75\%$ of the
primary Roche radius \cite{Sulkanen}, but this estimate is only
certain to within about $10\%$.  This uncertainty makes it
virtually impossible to estimate $\alpha_{\star}$ adequately for
use in Eq.\ (\ref{ComptonWavelength}) from present observational
data . Future observations of HeCV systems, either from advanced
ground-based instruments such as the Keck Interferometer, or from
space-based instruments such as the Space Interferometry Mission
\cite{SIM} and the Terrestrial Planet Finder \cite{TPF} could
allow direct measurement of $\alpha_{\star}$ by optically imaging
the detailed structure of close binary systems.

For cases when $\alpha_{\star}$ cannot be accurately determined,
the dependence of the Compton wavelength of the graviton on this
parameter (or on $\alpha_{\rm path}$) can be eliminated by subtraction
of two observations of the source.  Consider the situation shown
in Figure \ref{OrbitSubtraction}, where the gravitational and
electromagnetic signals are monitored when the Earth lies on one
side of its orbit, and again six months later, when it lies on
the other side of its orbit.

When the Earth is in position $1$, the phase difference between the
electromagnetic and gravitational wave signal can be written
\begin{equation}
   \Delta_{1} = {{\omega_{o}D}\over v_{g}} - \omega_{o}D + A \ .
   \label{Delta1}
\end{equation}
Similarly, when the Earth is in position $2$, the phase difference
may be written
\begin{equation}
   \Delta_{2} = {{\omega_{o}(D + L)}\over v_{g}} - \omega_{o}(D +
   L) + A \ ,
   \label{Delta2}
\end{equation}
where $L \leq 2 {\rm AU}$ is the path length difference for
the time of flight between the two measurement.  Subtraction
eliminates the unknown quantity $A$, yielding
\begin{equation}
   \Delta_{2} - \Delta_{1} = \omega_{o} L \left( {1 \over v_{g}} -
   1 \right) \ .
   \label{OrbitDifference}
\end{equation}
Defining the fractional change in phase from this quantity gives
\begin{equation}
    \epsilon = {{\Delta_{2} - \Delta_{1}} \over {\omega_{o} L}} = {1
    \over v_{g}} - 1 \ ,
    \label{FractionalOrbitalDifference}
\end{equation}
which in terms of the Compton wavelength becomes
\begin{equation}
    \lambda_{g} = {\pi \over \omega_{o}}\sqrt{{1 \over 2} \left(1 +
    {1 \over \epsilon} \right)} \ .
    \label{ComptonWavelength2}
\end{equation}
The unknown parameter, $\alpha$, has been eliminated from the 
expression for $\lambda_{g}$, but at the cost using a much shorter 
characteristic distance, $L \ll D$.  This approach amounts to 
measuring the phase lag between the two signals over the time it takes 
to cross the Earth's orbit (at most)\footnote{An improvement of 
roughly a factor of two may be obtained by combining the Earth's 
orbital motion with the proper motion of the binary system relative to 
the Sun, which will typically be of order a few AU per year}, as 
opposed to the time it takes to propagate over the Earth-source 
distance, leading to a great loss of precision in the measurement of 
$\lambda_{g}$.

\section{Obtaining a Mass Bound}\label{sec:AMCVn}

In order to estimate a bound on the graviton mass, assume a
null result for the measurement of the phase difference, $\Delta$,
between the two signals.  The size of $\Delta$ (and therefore
$\epsilon$) will then be limited only by the uncertainty in the
measurements of the phase.  Combining the uncertainty of the
gravitational phase measurements with the electromagnetic phase
measurements in quadrature yields
\begin{equation}
    \Delta \rightarrow \delta\Delta = \sqrt{\delta \phi_{em}^{2} +
    {1 \over 4} \delta \phi_{gw}^{2}} \ ,
    \label{Uncertainty}
\end{equation}
where $\delta \phi_{em}$ and $\delta \phi_{gw}$ are the uncertainties
in each of the phases.

For observations with a space-based interferometer, the error in
phase measurements can be estimated as the ratio between the
sampling time and the total integration time.  For LISA, the
sampling time is expected to be of order $1$ s, with total
integration times of $1$ yr = $3 \times 10^{7}$ s, yielding
$\delta\phi_{gw} = 3 \times 10^{-8}$. The CBA reports a $0.0004$s
uncertainty over the $1028.7325$s period of AM CVn, yielding a
phase uncertainty of $\delta\phi_{em} = 4 \times 10^{-7}$.

For the case of AM CVn, the value of $\alpha_{\star}$ is still not 
known, so bounds on the graviton mass must be derived from Eq.\ 
(\ref{ComptonWavelength2}).  As shown in Figure 
\ref{OrbitSubtraction}, the characteristic distance $L$ is simply the 
path length difference between the two measurements.  If the 
inclination of the binary system to the plane of the Earth's orbit is 
$\beta$, then the characteristic distance is $L = 2 \cos(\beta) {\rm 
AU}$.  For AM CVn, which lies at ecliptic latitude $\beta = 
37.4^{\circ}$, this yields $L = 2.38 \times 10^{11}$ m.  Using this 
value, Eq.\ (\ref{ComptonWavelength2}) gives a bound
\begin{equation}
    \lambda_{g} > 5 \times 10^{14}\ \mbox{m} = 5 \times 10^{11}\
    \mbox{km} \ ,
    \label{Bound1}
\end{equation}
or, in terms of the graviton mass, $m_{g} < 2 \times 10^{-21}$ eV, 
about a factor of five worse than the present bound based on the 
motion of Mars.  Even this weak bound would be of interest, however, 
since it is based on the {\it dynamics} of the gravitational field, 
({\it i.e.}, gravitational waves, rather than the static Yukawa 
modifications of the Newtonian potential).

If the value of $\alpha_{\star}$ could be determined precisely
({\it e.g.}, by monitoring ellipsoidal variations the light curve
in the infrared, as suggested in Section \ref{sec:IBWD}, or with
future optical interferometer observations), such that the
uncertainties $\delta\alpha_{\star} \lesssim 10^{-7}$, then Eq.\
(\ref{ComptonWavelength}) could be used to bound the graviton
mass. The distance to the known IBWD systems is typically of
order $D \simeq 100$ pc; combining this with a typical orbital
period of $P \simeq 1500 {\rm s}$ yields
a bound
\begin{equation}
    \lambda_{g} > 1 \times 10^{18}\ \mbox{m} = 1 \times 10^{15}\
    \mbox{km} \ ,
    \label{Bound2}
\end{equation}
or $m_{g} < 1 \times 10^{-24}$ eV. This potential bound would be
a factor of four hundred more stringent that the present solar
system based bound, and would be better than the bounds obtained
from inspiraling black holes proposed by Will \cite{WillGraviton}
for all but very large black holes.

During the next decade, as we await the launch of LISA, the optical 
astronomers may well succeed in further reducing the uncertainty in 
their phase measurements.  If the optical signal phase error is 
reduced in Eq.\ (\ref{Uncertainty}) to the point where the dominant 
source of error is the gravitational wave phase measurement, then the 
bound obtained above could be improved by about another factor of 
five, to $\lambda_g > 5 \times 10^{15} {\rm km}$ ($m_g < 2 \times 
10^{-25} {\rm eV}$).

\section{SUMMARY}\label{sec:Ending}
After the initial detection of gravitational waves the challenge will 
be for the field to evolve into a productive observational science 
which makes firm contact with astrophysics, complementing the broad 
base of electromagnetic observations already supporting that field.  
The experiment proposed here is particularly appealing because it 
entails observations of {\it known sources} by space-based detectors.  
The existence of IBWDs has been verified (as opposed to more 
speculative sources, such as binary black hole coalescence events), 
and such objects are currently under study by observational 
astronomers.  Detailed gravitational wave observations can begin 
almost as soon as a space-based interferometer such as LISA is online.

We have shown that reliable bounds on the mass of the graviton of
order $\lambda_{g} > 1 \times 10^{15}\ $ km could be obtained through
detailed observations of the interacting binary white dwarf star
systems such as AM CVn. With the combination of detailed studies
of such binary systems by optical interferometers and
gravitational wave observations, this could be a very robust
bound, several orders of magnitude greater than the current best
bounds from solar system observations.

If one assumes the graviton to be a massless particle, as
predicted by general relativity, then the same measurements
described here can be employed to determine the structure of the
binary star system. If the graviton is massless, then any phase
difference measured between the gravitational wave and
electromagnetic signal must be due to effects in the binary
system. Setting $v_{g} = 1$ in Eq.\ (\ref{FractionalLag}) yields
\begin{equation}
    \epsilon = {\Delta \over {\omega_{o}D}} = \alpha \ ,
    \label{AstroAlpha}
\end{equation}
showing that the difference in phase is simply an indicator of
the value of $\alpha = \alpha_{\rm path} + \alpha_{\star}$.  As was
shown in Section \ref{sec:AMCVn}, the value of $\alpha_{\rm path}$ is
expected to be negligible ($\alpha_{\rm path} = 7.9 \times 10^{-19}$
for AM CVn).  In the cases where $\alpha_{\rm path}$ can be ignored,
the measured phase difference will be a direct measure of the
value of $A$, which is the amount of phase by which the
electromagnetic signal leads the line of masses in the binary
system.  With good models of the matter stream from the
secondary Roche lobe overflow (such as the trajectories shown in
Figure \ref{RocheLobe}), a measurement such as this could allow
an accurate determination of the accretion disk radius and the
refinement of physical models for HeCV type stars solely from
gravitational wave observations.

\acknowledgements This work was supported in part by National
Science Foundation Grant No. PHY-9734834 and NASA Cooperative
Agreement No. NCC5-410.

\begin{table}
   \centering
   \caption{Properties of the six nearest interacting binary white dwarfs.}
   \begin{tabular}{lcccc}
    &Secondary Mass & Orbital Period & Distance & Strain {\it h}\\
    Name & ($M_{\odot}$) & (s) & (pc) & ($\times 10^{-22}$)\\
     \tableline
   & & & & \\
   AM CVn       & $0.044$     & $1028.73$ & $101$       & $5.27$ \\
   EC15330-1403 & $\sim\ 0.04$ & $1119$    & $165$       & $5.36$ \\
   CR Boo       & $\sim\ 0.03$ & $1491$    & $157$       & $2.82$ \\
   V803 Cen     & $\sim\ 0.03$ & $1611$    & $\sim 100$  & $4.20$ \\
   CP Eri       & $\sim\ 0.03$ & $1724$    & $\sim 100$  & $4.02$ \\
   GP Com       & $\sim\ 0.02$ & $2791.2$  & $165$       & $1.77$ \\
 \end{tabular}
 \label{BinaryTable}
\end{table}

\begin{figure}
  \caption{The background of close binaries in the Galaxy, plotted
  with the projected sensitivity of LISA. The six most
  well understood AM CVn type binaries are indicated.  The assumed
  bandwidth is $10^{-7}$ Hz.}
  \label{WDCurve}
\end{figure}

\begin{figure}
  \caption{Schematic of a binary star system which has observable
  electromagnetic and gravitational wave signals.  If the graviton is
  a massive particle, then the phase fronts of the gravitational
  signal will lag behind those of the electromagnetic signal.}
  \label{BinaryIdea}
\end{figure}

\begin{figure}
  \caption{The Roche lobe of the primary star is shown as a dark 
   line; matter overflows from the secondary star in a coherent matter 
   stream, shown as trajectories falling towards the white dwarf 
   (indicated by $\star$ at the center of the Roche lobe; figure 
   adapted from Flannery(1975) ).  Depending on the radius of the 
   accretion disk, the matter stream will strike the disk at some 
   angle $\theta$ which leads the binary axis; this angle is directly 
   proportional to the phase delay parameter $\alpha_{\star}$.  The 
   dashed circle represents an accretion disk radius which is $70\%$ 
   of the Roche radius.  A $10\%$ uncertainty in the radius of the 
   accretin disk will yield $\sim 5^{\circ}$ uncertainty in the value 
   of $\theta$.}
   \label{RocheLobe}
\end{figure}

\begin{figure}
  \caption{Schematic showing two observations of a source from opposite
  sides of the Earth's orbit.  The radius of the Earth's orbit is
  $r_{E}$ and the path length difference for signals propagating from
  a distant source is $L$.}
  \label{OrbitSubtraction}
\end{figure}

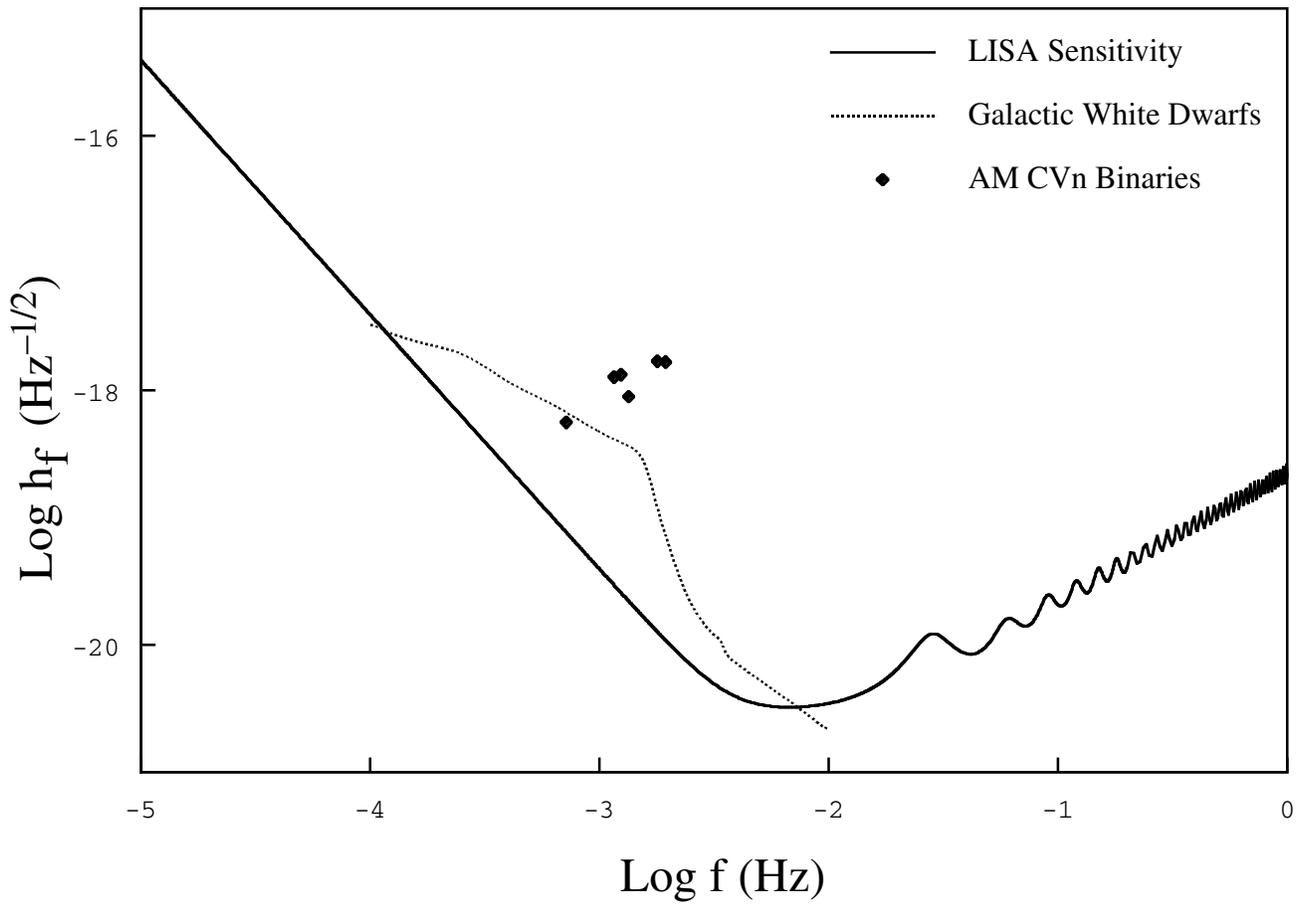
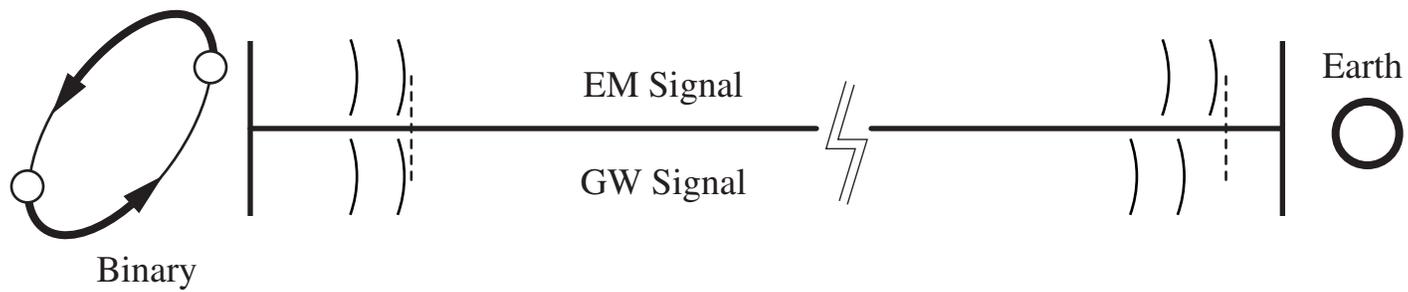
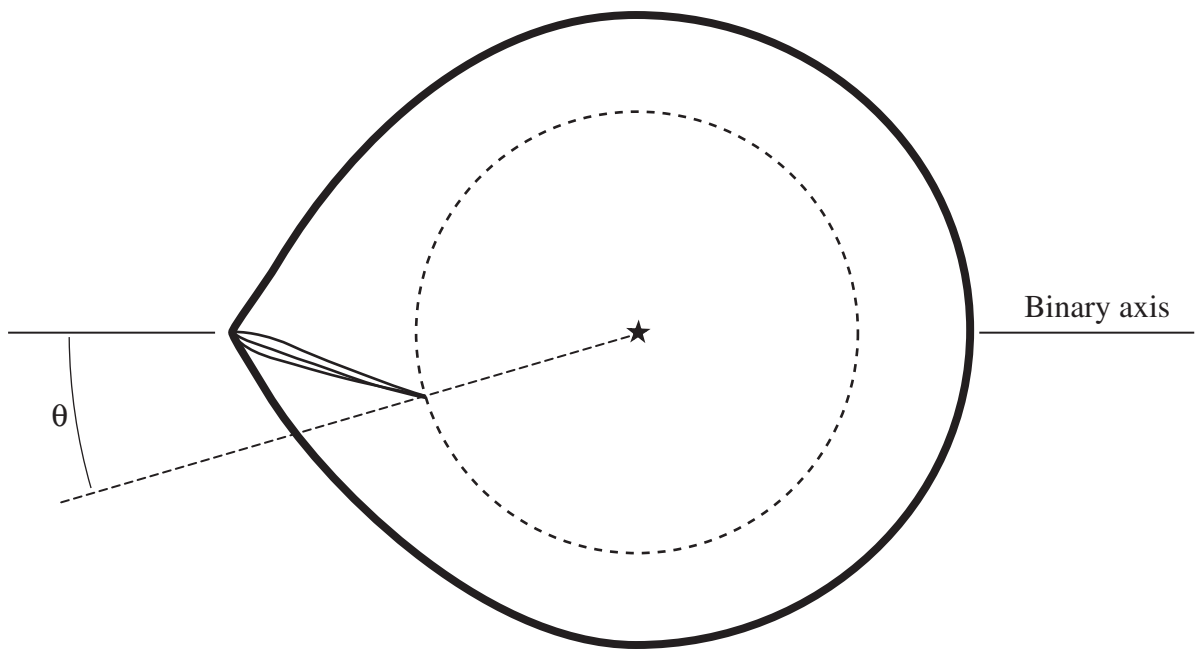
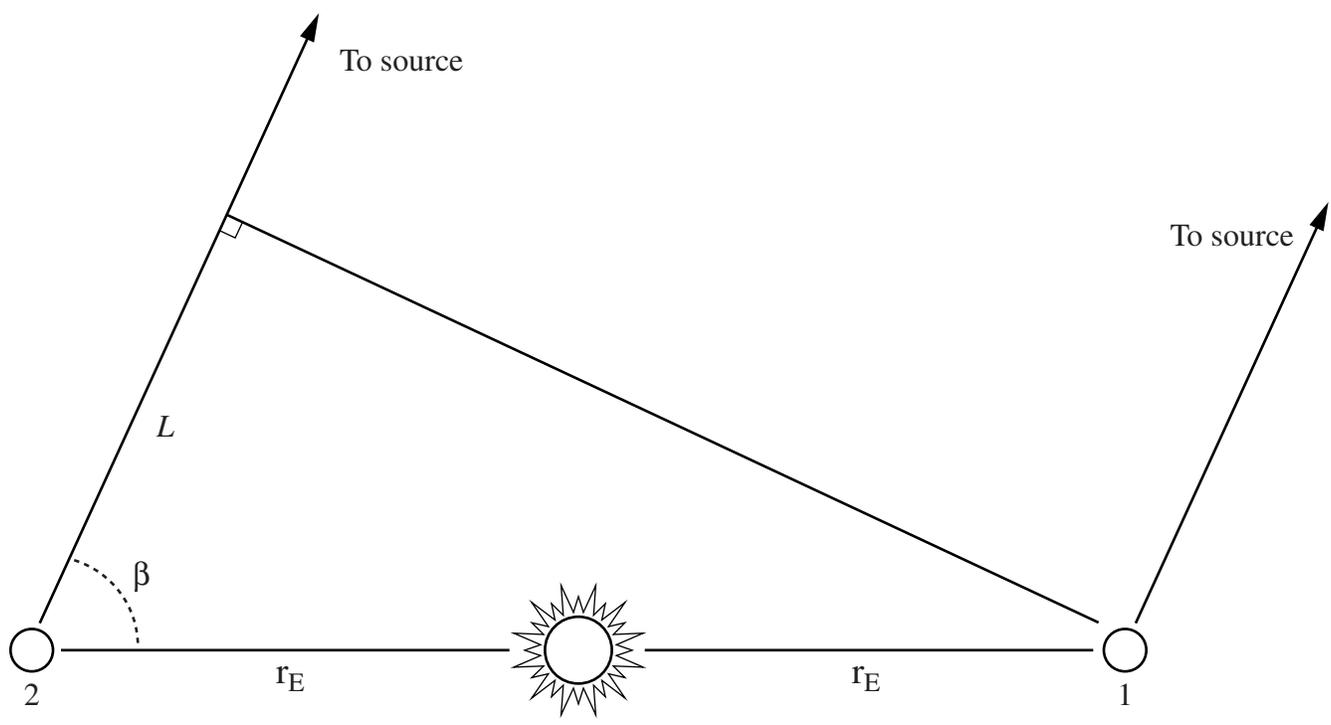
\end{document}